\documentclass[sigconf]{acmart}

\usepackage{url}
\usepackage{amsmath}
\usepackage{graphicx}
\usepackage{booktabs}
\usepackage{tabularx}
\usepackage{booktabs}
\usepackage{balance}

\begin{document}

\title{A Few Observations About State-Centric Online Propaganda}

\author{Jukka Ruohonen}
\affiliation{\institution{University of Turku, Finland}}
\email{juanruo@utu.fi}

\begin{abstract}
This paper presents a few observations about pro-Kremlin propaganda between 2015
and early 2021 with a dataset from the East Stratcom Task Force (ESTF), which is
affiliated with the European Union (EU) but working independently from
it. Instead of focusing on misinformation and disinformation, the observations
are motivated by classical propaganda research and the ongoing transformation of
media systems. According to the tentative results, (i)~the propaganda can be
assumed to target both domestic and foreign audiences. Of the countries and
regions discussed, (ii)~Russia, Ukraine, the United States, and within Europe,
Germany, Poland, and the EU have been the most frequently discussed. Also other
conflict regions such as Syria have often appeared in the propaganda. In terms
of longitudinal trends, however, (iii) most of these discussions have decreased
in volume after the digital tsunami in 2016, although the conflict in Ukraine
seems to have again increased the intensity of pro-Kremlin propaganda. Finally,
(iv)~the themes discussed align with state-centric war propaganda and conflict
zones, although also post-truth themes frequently appear; from conspiracy
theories via COVID-19 to fascism---anything goes, as is typical to propaganda.
\end{abstract}

% Generated at http://dl.acm.org/ccs.cfm
%



\maketitle

\section{Introduction}

This paper answers to recent calls \cite{Abhishek21, Stray19} to advance
research on institutional responses to propaganda by providing a few
quantitative observations and an accompanying discussion about the
countermeasures taken by the ESTF. It was set in 2015 by the European Union as a
response to Russia's propaganda activities during the early conflict in
Ukraine~\cite{EUCouncil15}. Thereafter, the main activity of the task force has
been to debunk propaganda, unintentional inaccuracies, and other information
disorders. While the ESTF still explicitly limits itself to pro-Kremlin
messages, it acknowledges that this ``\textit{does not necessarily imply,
  however, that a given outlet is linked to the Kremlin or editorially
  pro-Kremlin, or that it has intentionally sought to disinform}''
\cite{ESTF21}. The focus on Russian propaganda and the accompanying
acknowledgment are important for framing the paper's scope. The term
state-centric serves to clarify this focus. This term (or some variation
thereof) is often used in the Internet governance literature to describe
cooperation arrangements and actions by nation states and their alliances, as
opposed to arrangements and actions by non-state stakeholders, such as
companies, non-governmental organizations, and standardization
bodies~\cite{Ruohonen20JCP}. In the present context the term excludes propaganda
activities taken by domestic actors, such as political parties and interest
groups. Six addition points should be further taken into~account, as follows:

\begin{enumerate}
\itemsep 2pt
\item{Due to the metaphorical Hobbesian anarchy of international relations, and,
  by extension, the so-called cyber space \cite{Ruohonen20JCP}, at least all
  major powers can be reasonably assumed to participate in some propaganda
  activities in the cyber space~\cite{Thomas14}.}
\item{A response to state-sponsored propaganda is often state-sponsored
  counterpropaganda, and take note that ``\textit{propaganda against propaganda
    is just another propaganda}'' \cite{Lasswell35}. The risks associated with
  counterpropaganda are also well-recognized; it may easily turn against itself
  by altering the perceptions of domestic audiences, among other
  things~\cite{Hellman17}.}
\item{Analogously to offensive cyber (in)security operations, state-sponsored
  propaganda in the cyber space is carried out through different covert actions
  \cite{LandonMurray19}, and, therefore, like with cyber attacks, attribution of
  these activities is difficult.}
\item{State-sponsored propagandists---or their superiors, or both of them---are
  educated and trained well; they may be affiliated with armed forces or
  intelligence agencies, or both; they have sufficient resources and set
  strategic goals; and so forth.}
\item{The history of the ESTF has been closely related to the concept of hybrid
  warfare. The concept is difficult to define, but, in essence, it refers to the
  use of both conventional and unconventional, or kinetic and non-kinetic, means
  of warfare. The latter means include cyber attacks and propaganda, among other
  things. In this regard, there is a large body of academic literature on the
  different national perceptions of information security, information warfare,
  cyber security, cyber war, and related concepts. A common presumption in this
  literature is that Russian understanding of the terms differ from those used
  in Europe and the United States~\cite{Fridman17, Hellman17}.}
\item{Due to the earlier points, normative stances should be avoided in academic
  propaganda research. Regarding this paper, it should be acknowledged that the
  case studied refers to a state-sponsored organization explicitly designed to
  counter state-sponsored propaganda. In this sense, the few observations
  presented are biased; only one voice is being heard. Nor does the paper make
  claims about whose information is correct and for what reasons; what strategic
  goals are involved; how much resources are used; and so on. To summarize and
  \underline{underline}: the ESTF and its activities have raised controversies
  about alleged politicization and normative stances~\cite{Wagnsson18}, but the
  present paper takes no part in these broader debates.}
\end{enumerate}

Given these brief introductory remarks, the following research questions are
contemplated and tentatively answered:
\begin{enumerate}
\itemsep 2pt
\item{Given the ESTF's data, how much of the propaganda can be roughly estimated
  to be targeted for domestic consumption?}
\item{Recently, there have been claims and concerns that, in Europe, Russian
  propaganda mainly targets Germany~\cite{DW21a}. These provide a sufficient
  motivation for the second research question: how heavily different countries
  have been hit by Russian propaganda, as handled and attributed by the ESTF?}
\item{Given a tentative answer to the second question, have there been any
  longitudinal changes in the spatial dimension?}
\item{What themes are generally present in the propaganda?}
\end{enumerate}

There is not much to say about the structure. But to follow the conventions
nevertheless: Section~\ref{sec: background} briefly summarizes the research
background on propaganda. The empirical showdown occurs in Section~\ref{sec:
  results}. Results and other things are discussed in Section~\ref{sec:
  discussion}.

\section{Background and Related Work}\label{sec: background}

The golden age of academic propaganda research was during the Cold War. The
stimulus had been given earlier, however. In the United States, for instance, it
was the aftermath of the First World War that prompted a newly founded
interdisciplinary interest on propaganda research. Initially led by figures such
as sociologist Paul Lazarsfeld, the interest led to many well-funded,
methodologically oriented, and practice-focused research programs that were
later continued by other well-known figures, such as critical theorist Theodor
Adorno and communication scholar Harold Lasswell~\text{\cite{Abhishek21,
    Nietzel16}}. Lasswell's work in the 1930s also resulted a definition:

\begin{quote}
``\textit{Let us be clear about the meaning of propaganda. Propaganda may be
    defined as a technique of social control, or as a species of social
    movement.  As technique, it is the manipulation of collective attitudes by
    the use of significant symbols (words, pictures, tunes) rather than
    violence, bribery, or boycott.  Propaganda differs from the technique of
    pedagogy in that propaganda is concerned with attitudes of love and hate,
    while pedagogy is devoted to the transmission of skill.}''
    \cite[p.~189]{Lasswell35}
\end{quote}

\begin{figure}[th!b]
\centering \includegraphics[width=4cm, height=4cm,
  angle=-90]{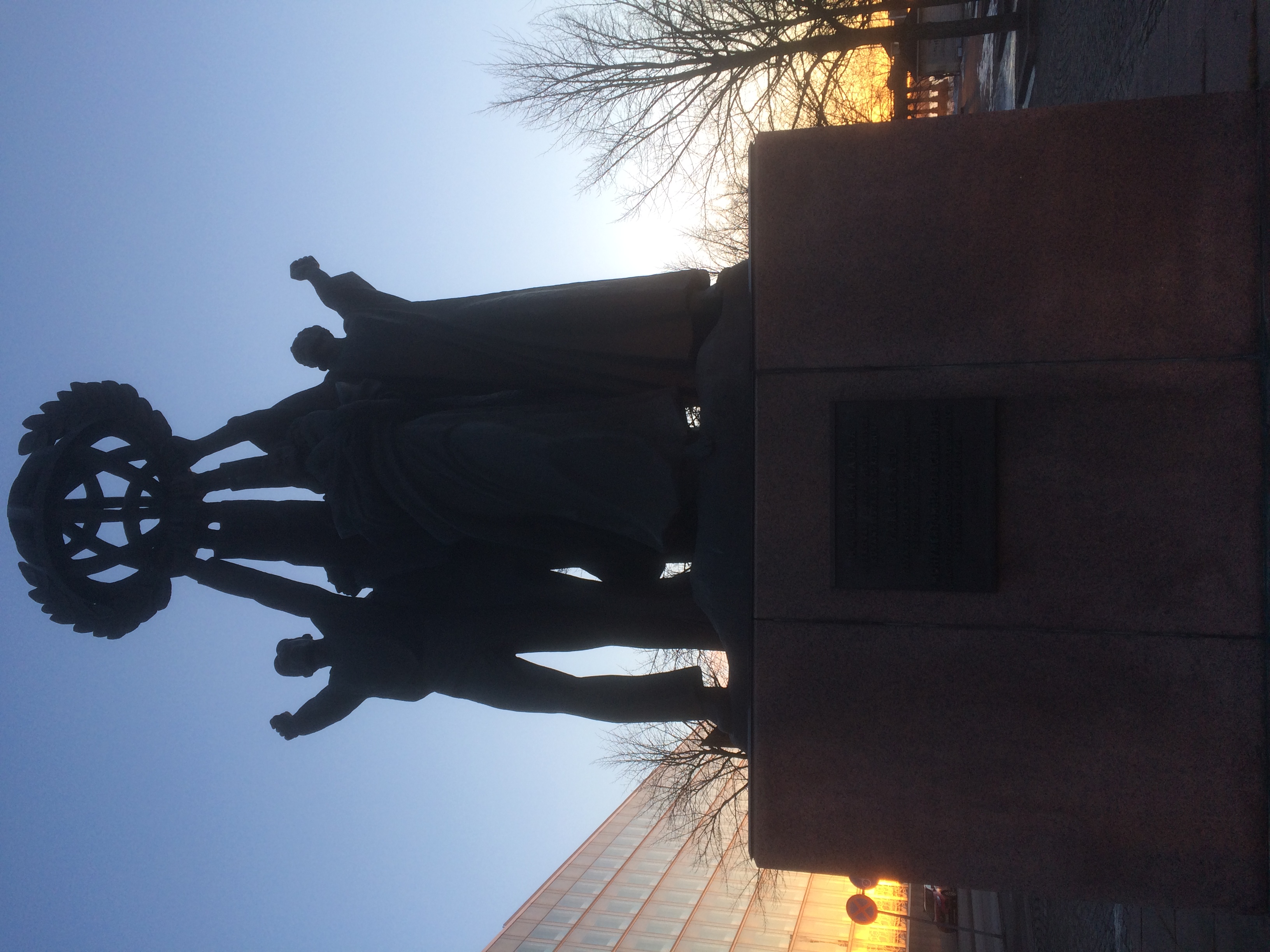}
\caption{``The Last Rites of Finnish Communism'' (a.k.a.~World Peace;
  Helsinki; Oleg Kirjuhin; author's personal collection)}
\label{fig: world peace}
\end{figure}

Here, the underlying emphasis of propaganda as a technique applies well to the
present day where ``\textit{significant symbols}'' are widely disseminated on
social media and other online platforms. But, of course, symbols have always
been an important part of propaganda. Consider the illustration in
Fig.~\ref{fig: world peace} as an example; it was donated to the City of
Helsinki in Finland by the City of Moscow about six months before the Berlin
Wall came down. The example is not by accident. When considering the history in
Europe, it was also a controversy over a sculpture in Estonia that, in 2007,
prompted a propaganda campaign and a set of state-sponsored cyber attacks,
which, in turn, later spurred the early cyber security initiatives in the
European Union~\cite{Ruohonen16GIQ}. Even moving the location of ``real-world''
symbols---let alone tearing them down---thus still carries relevance in
propaganda activities. Today, however, online symbols are what provide an
effective and cost-benefit-cheap propaganda alternative.

Propaganda research can be seen as a psychology of
ideology~\cite{Silverstein87}. Given widespread political polarization in
contemporary democracies, the underscoring of emotions---the ``\textit{attitudes
  of love and hate}''---indeed seems fitting for this particular
``\textit{technique of social control}''.  With these ingredients and an agency
of three actors (``\textit{the enemy, the ally, and the neutral}''), according
to Lasswell, the strategic goal of a propagandist is to intensify the attitudes
propagated, reverse any hostile attitudes, and attract those who remain
indifferent~\cite{Lasswell27}. Thus, the academic framework for propaganda
research was well-established already well-before the Second World War. The
academic propaganda research continued throughout the Cold War. But while the
amount of papers published on propaganda has continued to steadily grow, at some
point researchers started to perceive it merely as a historical
topic~\cite{Tal20}. The arrival of the so-called ``post-truth'' era in the 2010s
changed the perceptions.

Currently, propaganda is implicitly studied under the labels of misinformation
and disinformation. The former refers generally to unintentional adoption or
amplification of misleading information. Disinformation, in turn, is commonly
defined as a malicious use of ``\textit{false, inaccurate, or misleading
  information designed, presented and promoted to intentionally cause public
  harm or for profit}''~(\text{\cite[p.~2]{EU18a}}; for other definitions and
taxonomies see~\cite{Fetzer04} and \cite{Kapantai20}). Again, the definition
underlines both the techniques and the strategic goals. Thus, arguably, even
though seldom explicitly spelled, disinformation research is propaganda
research. Sure: it can be argued that disinformation differs from propaganda
because it tries to distort reality itself instead of relying on
persuasion~\cite{Borrell21, Cook20}, but, then again, it can be also argued that
distortion of reality is merely another form of
persuasion~\cite{vonMoltke16}. The intention to deceive is a distinct
characteristic of disinformation~\cite{Stewart21}, and intention is always
present also in propaganda. Actually, throughout the history, propaganda has
been cloaked with various alternative concepts, particularly by those involved
in propaganda activities~\cite{Bradley43, Miller15}. Although these
terminological nuances do not matter for the purposes of the few quantitative
observations presented, propaganda can be argued to be a better term than
disinformation due to the state-centric focus.

In addition to terminological similarities, there are methodological parallels
between ``classical'' propaganda research and the more recent disinformation
research. It was again Lasswell who, in the 1960s, pioneered content analysis,
defined broadly as ``\textit{systematic empirical studies of the messages
  transmitted in a process of communicacation}''~\cite[p.~57; original
  misspelling]{Lasswell68}. Today, content analysis of propaganda uses both
qualitative and quantitative methods. On the qualitative side, discourse
analysis has been a common way to analyze the significant symbols in propaganda
content~\cite{Recuero21}. On the quantitative side, topic modeling serves a
similar function for content analysis~\cite{Sear20}, as does---given that
propaganda is ``\textit{the politics of the heart}'' \cite{Thomas14}---sentiment
analysis~\cite{Ruohonen20MISDOOM}. But what is more fundamental and thus
interesting is Lasswell's classification of communication flows in a society
into a value-institutional framework.

This framework contains eight layers: \textit{power} (i.e., politics),
\textit{enlightenment} (e.g., science), \textit{wealth} (i.e., economics),
\textit{well-being} (e.g., health), \textit{skill} (e.g., education),
\textit{affection} (e.g., family), \textit{respect} (e.g., classes), and
\textit{rectitude} (e.g., ethics)~\cite{Lasswell68}. There are two reasons why
the framework is fundamental for propaganda research. First: the strategic goals
of propaganda differ between a layer and another layer. The paper's focus is on
power and politics, and, due to the state-centric focus, specifically in power
politics between states. But given the freedom of thought, it would be easy
enough to pick a different perspective and another layer. Propaganda against
enemies, in favor of allies, and for persuasion of indifferent; the poor and the
economy layer; tobacco, the pandemic, opioids, and the health layer; racism and
the respect layer; and what have you. Though, second, what remains difficult for
any propagandist, including a state-sponsored one, is that the layers evolve at
different~paces.

\begin{figure}[th!b]
\centering \includegraphics[width=5.2cm, height=4.9cm,
  angle=-90]{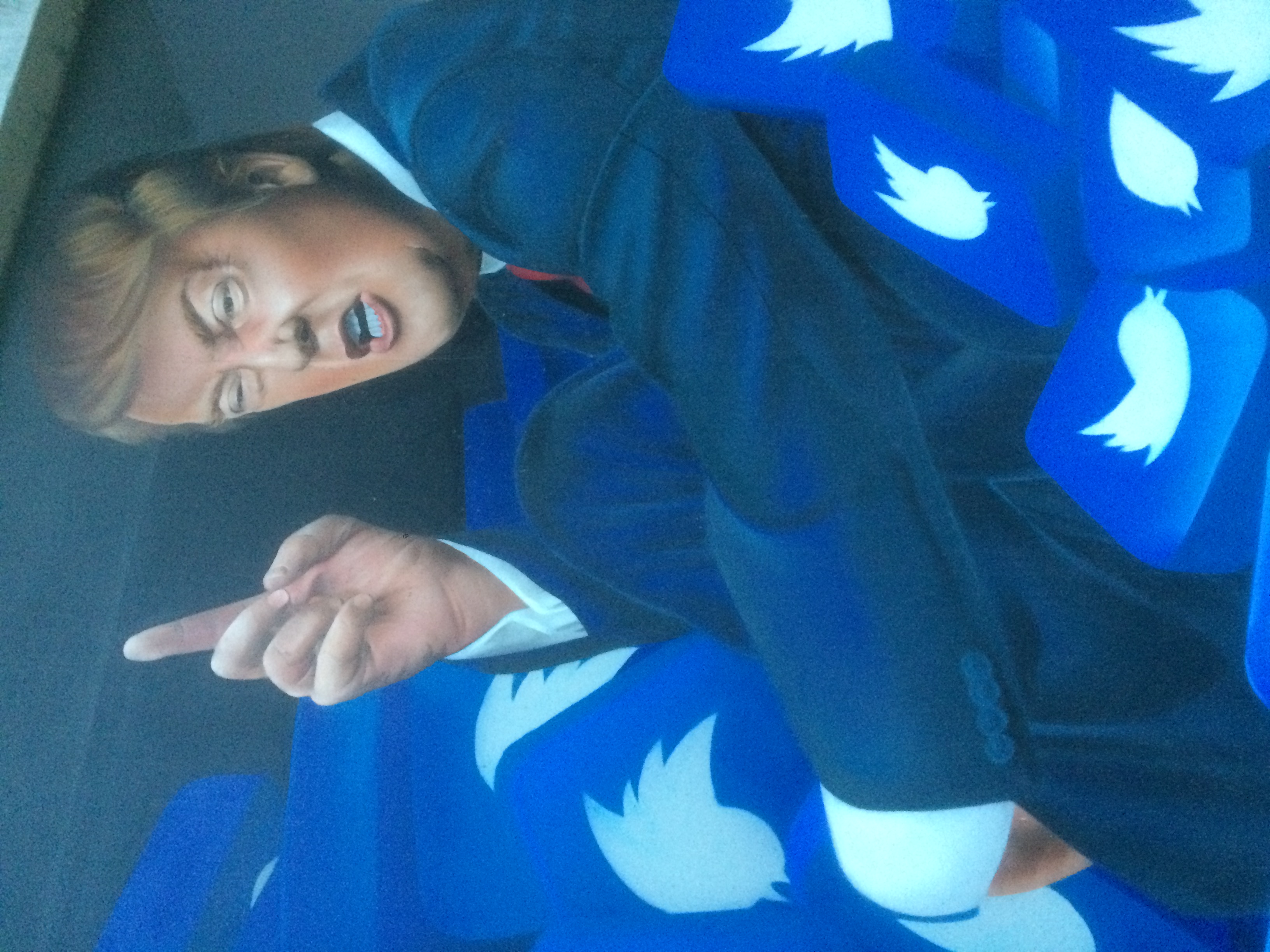}
\caption{``Riding on the Waves of the Digital Tsunami'' (Helsinki; unknown
  artist; author's personal collection)}
\label{fig: trump}
\end{figure}

A basic premise from sociology, institutional economics, and related fields
needs to be repeated; even changing the play of the game (i.e., at the power
layer) can take years, but changing things at the rectitude layer takes hundreds
of years or more~\cite{Williamson00}. In other words, values, norms, and
institutions change only slowly. For a propagandist they are often lamentably
resilient. But, eventually, changes at the power layer drift onto the other
layers. Currently, it is the tsunami of digital information and epistemic
inequality that is changing communication that is changing
institutions~\cite{Gurri20, Zuboff20a}. Like many politicians
(cf.~Fig.~\ref{fig: trump}), state-sponsored propaganda largely rides on the
waves caused by the tsunami. These waves have prompted numerous ideas about
countermeasures beyond counterpropaganda. The literature offers many options;
these range from the so-called whole society approaches, which may include
states' intelligence apparatuses~\cite{Ivan21}, media literacy and literature
skills~\cite{Ruohonen21ICEDEG}, fact-checking~\cite{Young18},
regulation~\cite{Borrell21, EC18b}, collaboration between governmental agencies
and online platforms~\cite{HybridCoE21}, sustainable business models for
journalism~\cite{Ruohonen20JMCS}, and other things, to various semi-manual or
fully computational solutions, including classification~\cite{Kausar20}, content
moderation~\cite{Stewart21}, network analysis~\cite{Starbird19}, and many other
things. But do these fix the problem? Time will tell, but the current foresight
from academia and elsewhere seems pessimistic. Why? The current interaction
between the distinct value-institution layers is a major source feeding the
pessimism.

In particular, the dependencies and interactions between the power,
enlightenment, wealth, and skill layers constitute a big problem, the core of
which is in the information tsunami's relation to media systems. To better
understand this kernel, one should ask a simple but fundamental question: What
is the antithesis of propaganda? For Lasswell it was
deliberation~\cite{Lasswell27}, and it is deliberation against which the
tsunami's waves have hit hard. Deliberative democracy requires a
well-functioning media system. However, the political economy of
propaganda---the interactions between media systems and political systems---has
largely been neglected in recent research~\cite{Abhishek21}. A historical
perspective is again needed; democracies have been in a similar situation
several times after the invention of radio and public broadcasting in the 1930s,
but national responses have diverged each time~\cite{Fukuyama20}. Here, the
1930s provides a good parallel for the present day---yet not because of the
political events but because of the historical transformation of media. The
deteriorating trust in media in the face of propagandists all over the world was
a grave concern back then~\cite{Berchtold34}. Today, there something eerie to
read melancholic memoirs about that time, such as the one that ends in:

\begin{quote}
``\textit{But I do urge that these possibilities exist, and that those who care
    for literature might turn their minds more often to this much-despised
    medium, whose powers for good have perhaps been obscured by the voices of
    Professor Joad and Doctor Goebbels.}'' \cite{Orwell46}
\end{quote}

Finally, it should be emphasized that the ESTF's early efforts were largely
related to countering war propaganda. As any analysis of any war should
presumably reveal, it is precisely war that makes state-centric propaganda shine
its bleakest luminescence. War propaganda is almost always both directed to home
consumption and manufactured for export~\cite{Berchtold34}. For the former
audience, the persuasion of those indifferent is a common goal; here, the words
hate and enemy often manifest themselves concretely. As the history again
eagerly testifies, sometimes wartime propaganda escalates into extreme measures
involving humiliation, dehumanization, and even
torture~\cite{Miller15}. Regardless of the measures taken, as said, war
propaganda is usually carried out by well-resourced and well-educated
specialists. But often also rank-and-file amateurs take part with solemn but
almost poetical propaganda of their own: ``\textit{Those who kill for pleasure
  are sadists \// Those who kill for money are mercenaries \// Those who kill
  for both are RECON \// WE DEAL IN DEATH}'' \cite{Friedman14}.

\section{Results}\label{sec: results}

The dataset is based on the propaganda cases handled and countered by the
ESTF. These were simply retrieved from the task force's website in 6 April,
2021. In total, $n = 11,397$ propaganda cases are present, spanning a period
from January 2015 to March 2021. The required quantification was done from the
meta-data associated by the ESTF with these cases. As there is not much more to
add about the dataset, its collection, and its processing, the dissemination of
the results can proceed immediately. One, two, three, launch.

\begin{figure}[th!b]
\centering \includegraphics[width=\linewidth, height=8cm]{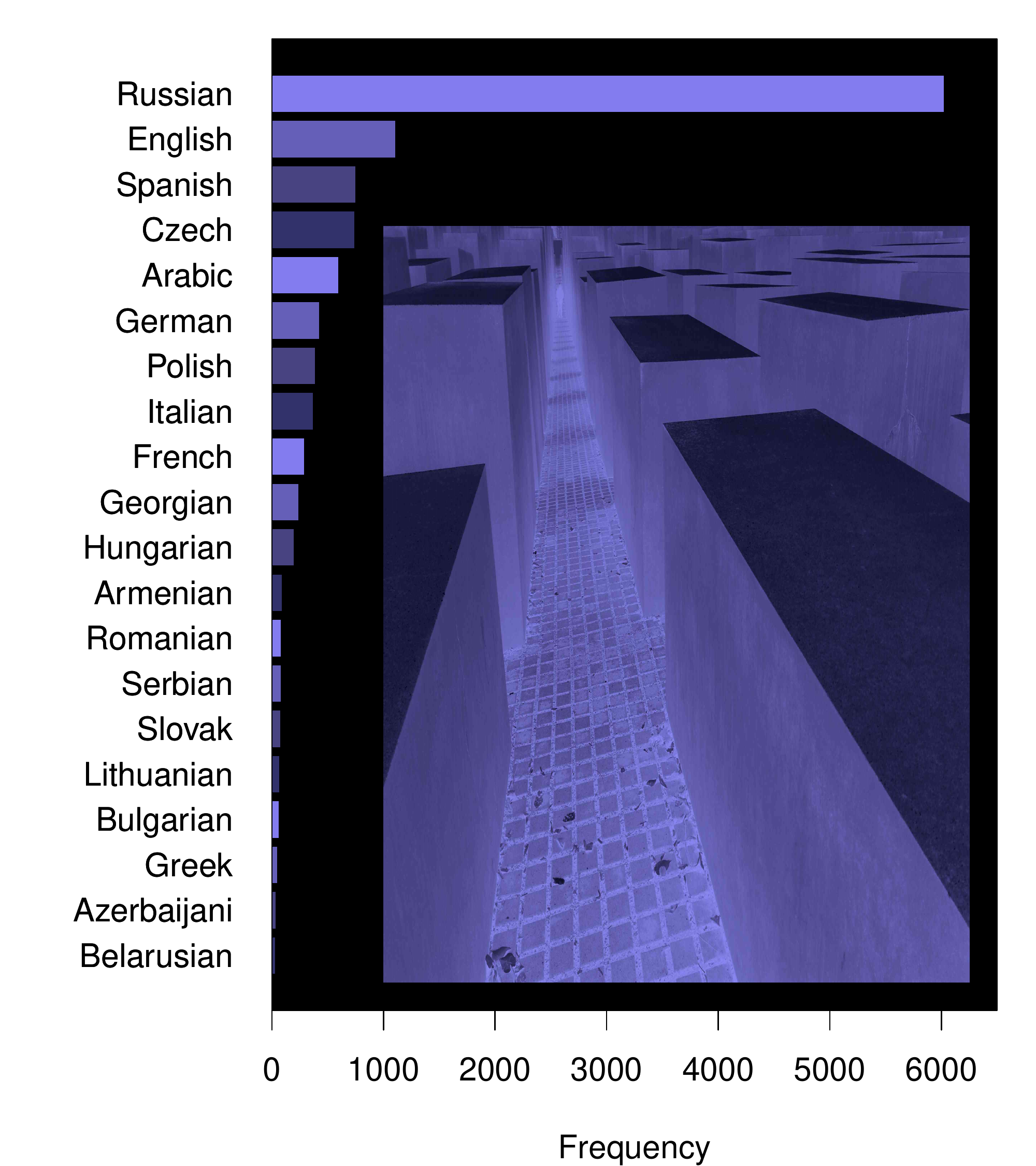}
\caption{Top-20 Languages Used in the Propaganda}
\label{fig: lang}
\end{figure}

One, Fig.~\ref{fig: lang} shows the most frequent languages used in the
propaganda accounted by the ESTF. Most of it was in Russian, which, partially,
supports an assumption of domestic targets. Through, it should be underlined
that pro-Kremlin propaganda often specifically targets Russian-speaking
minorities in Europe. For instance, about 34\% of Latvians, 30\% of Ukrainians
and Estonians, and 8\% of Lithuanians speak Russian at home~\cite{CIA21}. The
countries associated have also been under a specific radar by the ESTF and other
related organizations~\cite{Saurwein20}. Although English was the second most
frequent language, it is interesting to note the relatively large amounts of
propaganda written in Spanish, Czech, and Arabic. Indirectly, this observation
aligns with the recent arguments that academic propaganda research has had a too
narrow attention, mainly concentrating on the United States and, perhaps to a
lesser extent, Europe~\cite{Abhishek21}.

\begin{figure}[th!b]
\centering \includegraphics[width=\linewidth, height=5.5cm]{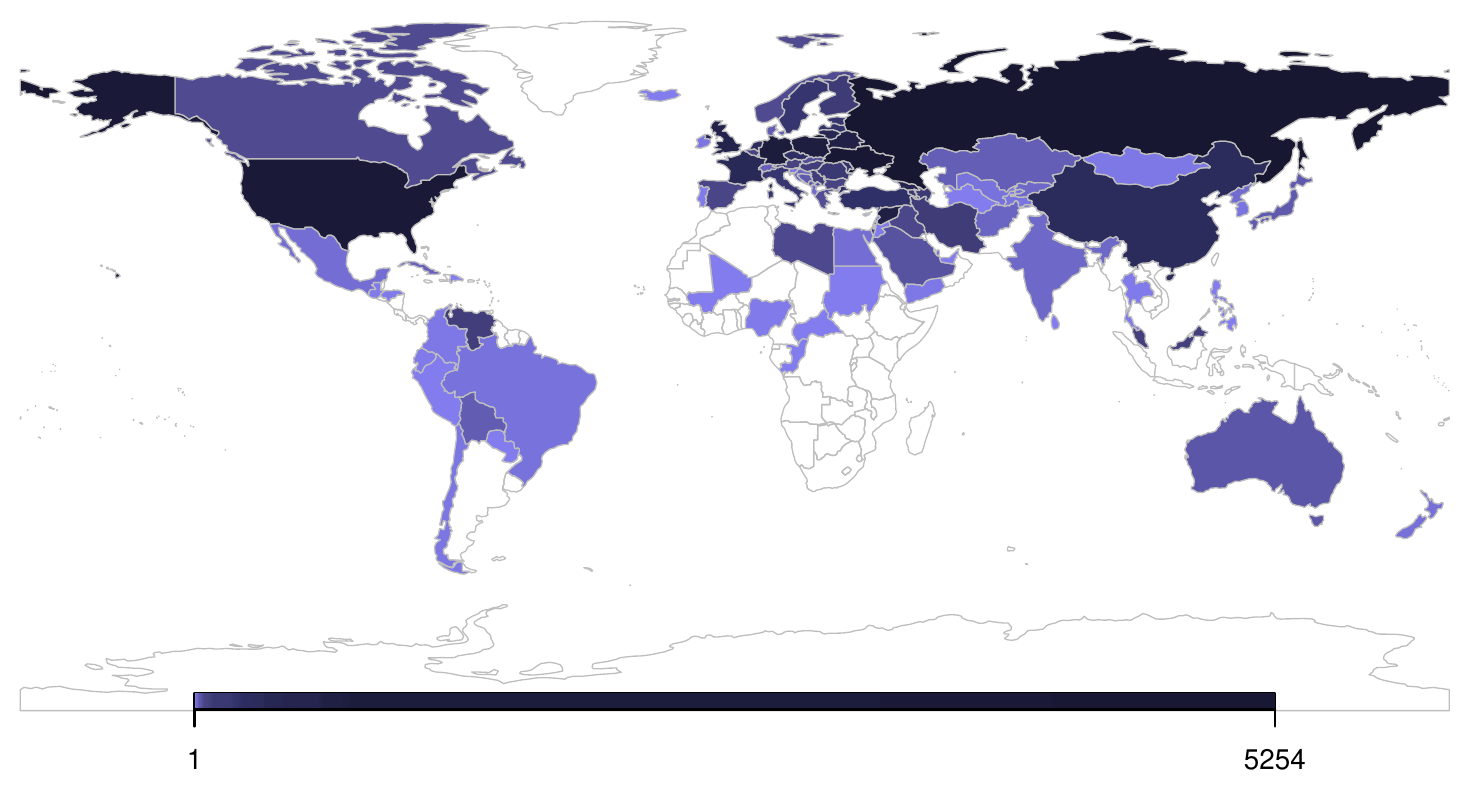}
\caption{Countries Discussed in the Propaganda}
\label{fig: countries}
\end{figure}

\begin{figure}[th!b]
\centering \includegraphics[width=\linewidth, height=5cm]{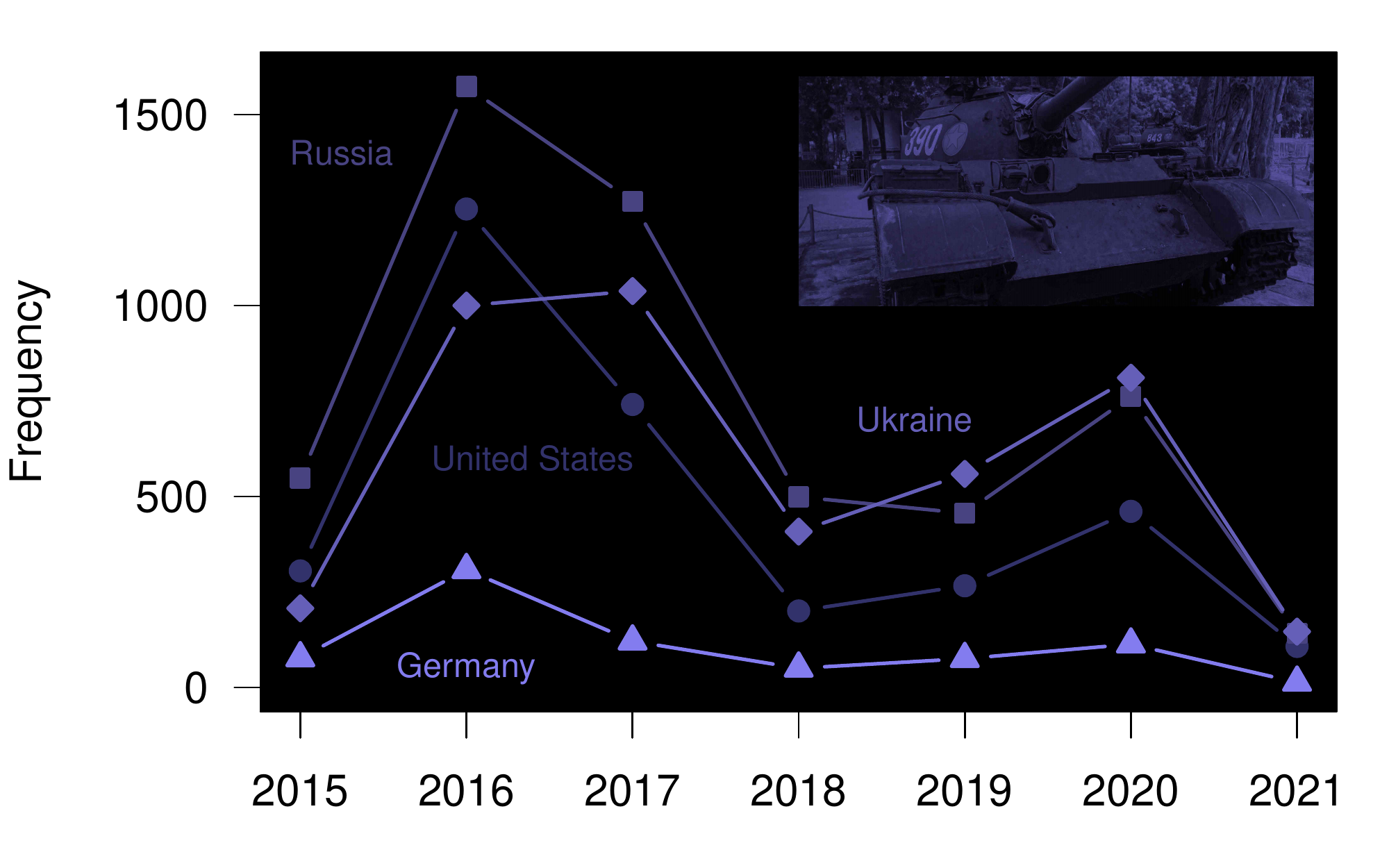}
\caption{Annual Trends of Four Countries Discussed}
\label{fig: trends}
\end{figure}

\begin{figure}[th!b]
\centering \includegraphics[width=\linewidth, height=9.4cm]{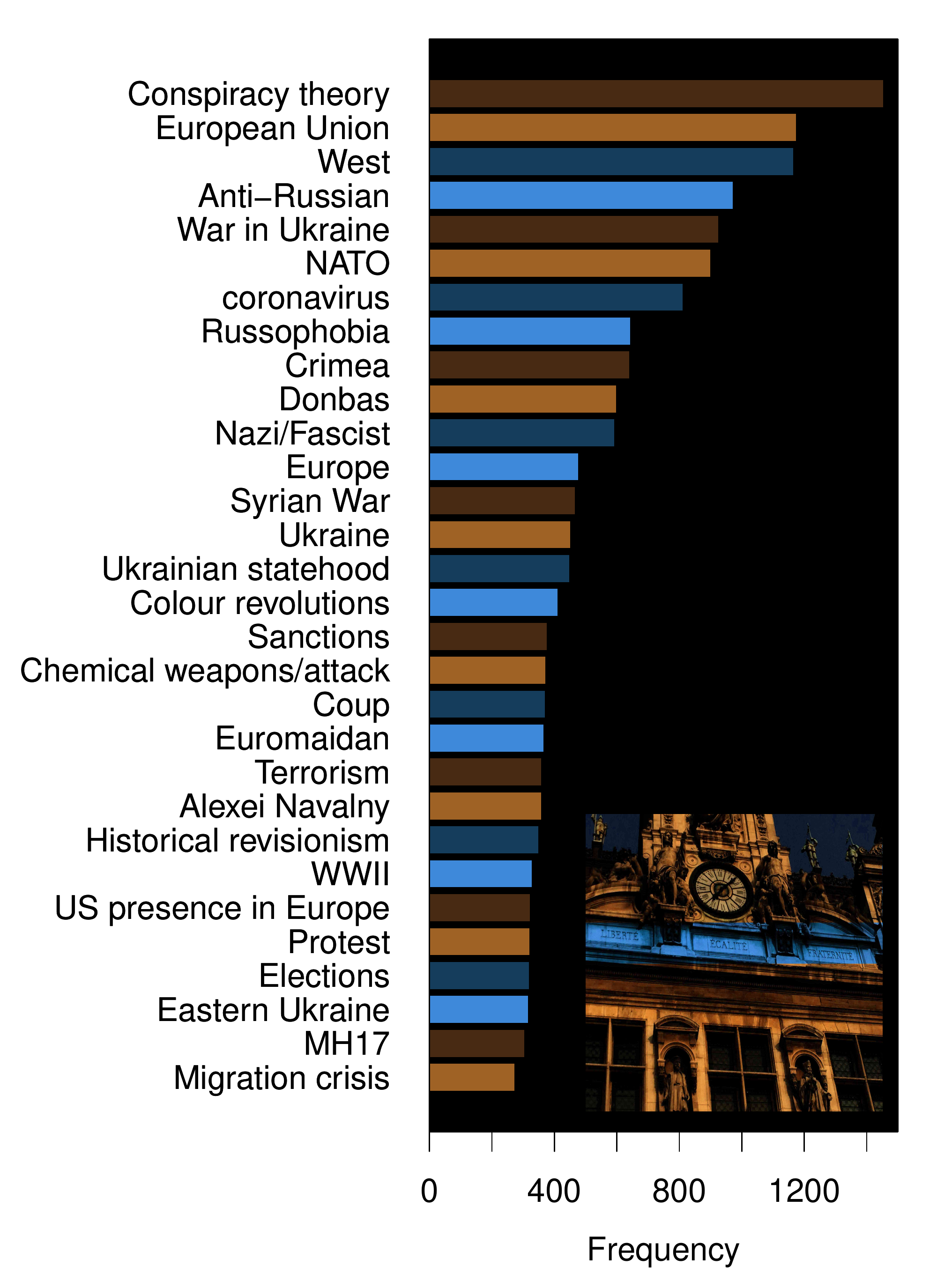}
\caption{Top-30 Themes Discussed in the Propaganda}
\label{fig: themes}
\end{figure}

Two, the illustration in Fig.~\ref{fig: countries} further indicates that much
of the pro-Kremlin propaganda has discussed Russia according to the ESTF's
data. This observation again hints about domestic as well as foreign
audiences. After Russia, the most discussed countries were Ukraine and the
United States, respectively. Within Europe, Germany, Poland, and the EU were the
most frequent targets of discussion, although, as can be seen, all European
countries have been discussed in the propaganda. A further noteworthy
observation from the figure is the relatively common mentions of other conflict
regions, including Georgia and Syria in particular. In terms of longitudinal
trends, most of the per-region mentions peak around 2016 during which the
tsunami's tidal waves first truly hit the shores. As can be further concluded
from the illustration in Fig.~\ref{fig: trends}, the surge has since decreased
in its intensity. However, the increasing hostility in the Russia-Ukraine
relations has also increased the propaganda references to these two countries
after circa late-2018. Although only early 2021 is covered in the dataset, the
longer annual trends indicate no particular increase in propaganda discussing
Germany.

Three, the breakdown of the most common keywords in Fig.~\ref{fig: themes}
reinforces the point about the propaganda's emphasis on conflict zones, defense,
and military affairs. Likewise, the earlier points about propaganda and
counterpropaganda are supported by the frequent addressing of anti-Russian
sentiments, whereas keywords such as conspiracy theories, COVID-19, extremism,
and historical revisionism generally coincide with the post-truth
topic. Although many states supposedly use these post-truth themes in their
propaganda, there are some nuances; among these is historical revisionism that
has been typical to Russian neoconservatism~\cite{VazquezLinan17}. Furthermore,
as could be expected, the themes vary from a country to another. Take Finland as
an example: the most frequent topics are the NATO, historical revisionism, the
Second World War, and the EU, although there are also mentions of some populist
Finnish politicians, journalists~\cite{Aro16}, and associated state-sponsored
propagandists. Take the theme of conspiracy theories as another example: here,
the most frequent countries discussed are the United States (by a large margin),
Russia, Ukraine, Belarus, and the United Kingdom, in the order of
listing---which does not correlate with the overall ranking of countries and
regions discussed. Each to their own.

\section{Discussion}\label{sec: discussion}

So what is sensible to say based on the observations presented? Some things seem
clear. Among these is the digital tsunami that caused the waves on which most
current propaganda surfs, whether state-sponsored or something else. Another is
the misconception that things like conspiracy theories would be merely
organically spreading misinformation originating from the fringes of the cyber
space. Instead, some of them belong also to the conventional toolbox of
well-educated, state-sponsored propagandists. If pedagogy is the transmission of
skill, as it was for Lasswell, a basic lesson about human emotions, among them
love and hate, would thus be the first advice for a future propagandist, as it
would have been in the~1930s.

What else? As was expected, much of the pro-Kremlin propaganda, according to the
task force's data, seems to be targeted for domestic consumption as much as
produced for export. There are no reasons to expect that propaganda from other
capitals would be much different. Otherwise, sure, there are differences;
according to the ESTF, Russian propaganda is largely about war propaganda
addressing both ``hot'' and ``cold'' conflict zones. This observation supports
the notions about hybrid warfare, which, however, is nothing novel from a
historical perspective. What is new is that what is old is often forgotten,
misunderstood, or manipulated; or perhaps this is old as well. But in terms of
international relations, what is historically new is the world's inevitable
interconnectedness also in terms of information in the open Internet. Data,
information, knowledge, and propaganda; it does not seem to matter. When one
country becomes a hotbed for one, other countries tend to follow. When it is
propaganda, they tend to also share the same hangover.

Generalizability of the observations presented is the most notable
limitation. In other words, it is difficult to say whether these truly reflect
Russian propaganda in general. As said, the ESTF itself supposedly has a certain
barycenter of its own, but, more generally, sampling is a known problem in
propaganda research~\cite{Silverstein87}. Another known problem is an
impulse-response type of an analysis~\cite{Starbird19}. The same applies to the
case at hand; it is difficult to say anything about how successful the task
force has been. Furthermore, the ESTF's current reactive whack-a-mole model
makes evaluation difficult, including any assessment over whether a more
proactive model would be plausible~\cite{Bjola17}. It can be left as a further
exercise to contemplate what a proactive model might look like---presently, it
suffices to note that the whole fact-checking paradigm may be doomed. To
understand why, one should return to the earlier definition and its emphasis of
social control.  To gain such control, epistemology is not a concern; anything
goes for a propagandist as long as it helps to advance the cause of control
through the public opinion~\cite{Wimberly17}. Another point is that the waves
from the tsunami are tidal waves against which any debunking effort runs short
on time and resources. In this regard, sufficient resourcing and staffing for
the ESTF have also raised critical questions previously~\cite{Stray19,
  Wagnsson18}. So, given these limitations, the earlier discussion with the
literature, and the observations presented, is there a hope for an improvement?
A tentative answer from the parallel 1930s seems again appropriate:

\begin{quote}
``\textit{There are no indications to encourage a hope that this propaganda war between nations will cease; there is every reason to believe that it will become more intense. It is a vicious game at which nations can play only by poisoning the minds of each other's nationals}.'' \text{\cite[p.~430]{Berchtold34}}
\end{quote}

Finally: As interesting the results and the ESTF case may or may not be, there
is a more important question at hand. To motivate it, consider again the two
Figures~\ref{fig: world peace} and ~\ref{fig: trump}. Again: neither one has
been chosen by accident. Pause for a moment to think: Why? Dot dot dot: for many
thinkers, the first figure symbolizes the ending of an era. It symbolizes the
end of the Cold War and the bipolar world order. In more polemical terms, it
symbolizes the triumph of liberal democracy over communism during \textit{that}
era (Fukuyama). In still polemical terms, it symbolizes the end of the
\textit{short} but extreme twentieth century~(Hobsbawm). But whichever polemic
one chooses, one still arrives to the latter figure. There is even a
\textit{shorter} time period in-between what the figures symbolize. Now, it was
argued in Section~\ref{sec: background} that values, norms, and institutions are
resilient. Then: How can it be that so much has changed during this shorter
period than the ``short'' previous century? Technology and the information
tsunami were the answers implicitly contemplated in this paper. But while these
may explain today's enemies, allies, neutrals, loves, and hates---the propaganda
of the early 21st digital century, the explanation still seems partial. To
proceed, it seems sensible that all eight value-institution layers discussed
should be~addressed.

\clearpage
\pagebreak
%\balance
\bibliographystyle{abbrv}
%\bibliography{eu}

\end{document}